\documentclass[aps,prl,twocolumn,preprintnumbers,showpacs,floatfix,nofootinbib]{revtex4-1}

\usepackage{mathtools}
\usepackage{bm}
\usepackage{siunitx}
\usepackage{caption,subcaption}
\usepackage{xcolor}
\usepackage[colorlinks=true,citecolor=blue,urlcolor=blue]{hyperref}
\usepackage{cleveref}

\let\originalleft\left
\let\originalright\right
\renewcommand*{\left}{\mathopen{}\mathclose\bgroup\originalleft}
\renewcommand*{\right}{\aftergroup\egroup\originalright}
\DeclarePairedDelimiter\abs{\lvert}{\rvert}

\newcommand*{\diff}{\mathop{}\!\mathrm{d}}
\newcommand*{\V}[1]{\ensuremath\bm{#1}}

\newcommand*{\cL}{\mathcal{L}}

\usepackage[normalem]{ulem}

\usepackage{rotating}

\newcommand*{\jewel}{\textsc{Jewel}}
\newcommand*{\pythia}{\textsc{Pythia}}
\newcommand*{\demix}{\textsc{Demix}}
\newcommand*{\fastjet}{\textsc{FastJet}}
\newcommand*{\emcee}{\textsf{emcee}}
\newcommand*{\SN}{\,\text{SN}}

\newcommand*{\solved}[1]{}

\newcommand{\Refc}[1]{Ref.~\cite{#1}}
\newcommand{\Refsc}[1]{Refs.~\cite{#1}}
\newcommand{\Reflc}[1]{Reference~\cite{#1}}

\newcommand{\nSD}{\ensuremath n_\text{SD}}

\begin{document}

\preprint{MIT--CTP 5219}

\title{Data-driven quark and gluon jet modification in heavy-ion collisions}
\author{Jasmine Brewer}
\email{jtbrewer@mit.edu}
\author{Jesse Thaler}
\email{jthaler@mit.edu}
\author{Andrew P. Turner}
\email{apturner@mit.edu}
\affiliation{
Center for Theoretical Physics, Department of Physics, Massachusetts Institute of Technology, 77 Massachusetts Avenue, Cambridge, MA 02139, USA
}


\begin{abstract}
Whether quark- and gluon-initiated jets are modified differently by the quark--gluon plasma produced in heavy-ion collisions is a long-standing question that has thus far eluded a definitive experimental answer.
A crucial complication for quark--gluon discrimination in both proton--proton and heavy-ion collisions is that all measurements necessarily average over the (unknown) quark--gluon composition of a jet sample.
In the heavy-ion context, the simultaneous modification of both the fractions and substructure of quark and gluon jets by the quark--gluon plasma further obscures the interpretation.
Here, we demonstrate a fully data-driven method for separating quark and gluon contributions to jet observables using a statistical technique called topic modeling.
Assuming that jet distributions are a mixture of underlying ``quark-like'' and ``gluon-like'' distributions, we show how to extract quark and gluon jet fractions and constituent multiplicity distributions as a function of the jet transverse momentum.
This proof-of-concept study is based on proton--proton and heavy-ion collision events from the Monte Carlo event generator \jewel\ with statistics accessible in Run 4 of the Large Hadron Collider.
These results suggest the potential for an experimental determination of quark and gluon jet modifications.
\end{abstract}

\maketitle


High-energy collisions between large nuclei at the Relativistic Heavy Ion Collider (RHIC) and the Large Hadron Collider (LHC) are a critical laboratory for studying the deconfined phase of QCD matter, the quark--gluon plasma, created in these collisions.
Collimated sprays of high-momentum hadrons, called jets, are produced copiously in these collisions and provide an important probe of the quark--gluon plasma they pass through.

A long-standing question is how the quark--gluon plasma resolves the color charge of high-energy QCD partons~\cite{CaronHuot:2008ni,Spousta:2015fca,Chien:2018dfn, Mehtar-Tani:2018zba,Qiu:2019sfj,Apolinario:2020nyw}.
Since jets can originate from either a quark or gluon, and subsequently carry information about their respective total color charge, it is crucial to understand differences in the energy loss and modification of these two categories of jets.
Unfortunately, accessing independent information about quark and gluon jets experimentally is very challenging because all jet measurements involve a mixture of contributions from both.

In this letter, we demonstrate a data-driven method to estimate both the quark and gluon jet fractions and their separate substructure modification in heavy-ion collisions.
Our method is based on a statistical technique called topic modeling, which was pioneered for applications to quark and gluon jet separation in proton--proton collisions in \Refsc{Metodiev:2018ftz,Komiske:2018vkc} and has been applied experimentally in \Refc{Aad:2019onw}.
We present a proof-of-concept that an extension of that technique can be used to extract differences in the modification of quark and gluon jets in heavy-ion collisions with the statistics anticipated in Run 4 of the LHC.
This is a critical step toward a model-independent determination of quark and gluon jet modification in heavy-ion collisions, which would have dramatic consequences for understanding the microscopic structure of the quark--gluon plasma.

A similar type of analysis was recently performed in \Refc{Sirunyan:2020qvi}, which used a measurement of the jet charge and templates for the jet charge distributions of quark and gluon jets to extract the gluon fraction in proton--proton and heavy-ion collisions.
In that study, the same Monte Carlo (MC) distributions were used as templates in both proton--proton and heavy-ion collisions, which makes the implicit assumption that the jet charge distributions of quark and gluon jets are unmodified by the quark--gluon plasma.
Here, we present a method that does not require templates and does not assume that substructure observables are unmodified by the plasma, allowing for simultaneous estimates of the modification of quark and gluon jet fractions and of their distributions.

Our method is based on a statistical technique called \demix~\cite{KatzEtAlDemix} that separates a pair of mixed probability distributions into two common underlying base distributions.
This method was demonstrated in \Refsc{Metodiev:2018ftz,Komiske:2018vkc} as a way to obtain excellent proxies for quark and gluon jets in proton--proton collisions.
Consider two probability distributions $p_1(x), p_2(x)$ for a jet observable $x$ that are a distinct mixture of the same two underlying base probability distributions $b_1(x), b_2(x)$.
Namely, we can express the mixture distributions as $p_j(x) = f_j \, b_1(x) + (1 - f_j) \, b_2(x)$, for distinct fractions $f_j$.
This expression of the $p_j$ is always ambiguous, however, since there are infinitely many ways to mix the base distributions $b_i(x)$ among themselves to obtain new distributions $\tilde{b}_i(x)$ from which the mixture distributions can be expressed as $p_j(x) = \tilde{f}_j \, \tilde{b}_1(x) + (1 - \tilde{f}_j) \, \tilde{b}_2(x)$ with new fractions $\tilde{f}_j$.
The idea behind \demix\ is to resolve this ambiguity by further requiring that the base distributions are \emph{mutually irreducible}~\cite{blanchard2016}.
Qualitatively, this means that neither base distribution contains any component of the other; a precise definition will follow shortly.
We refer to the mutually irreducible base distributions as \emph{topics}, for their relation to the broader field of topic modeling established in \Refc{Metodiev:2018ftz}.
See \Refsc{Dillon:2019cqt,Alvarez:2019knh,Dillon:2020quc} for other uses of topic modeling techniques in collider physics.

The notion of ``quark- and gluon-initiated jets'' is not well-defined at the hadron level.
Even at the level of a MC generator, where parton information from the hard process is available, there is still an ambiguity about how to associate final-state jets with their initiating parton.
Therefore, the quark and gluon topics we discuss in this letter do not correspond directly to any parton-level intuition about quark- and gluon-initiated jets~\cite{Gras:2017jty}.
Instead, these topics correspond to the \emph{operational definition} of jet categories introduced in \Refc{Komiske:2018vkc}, which defines the quark and gluon categories as the mutually irreducible (i.e., maximally separable) distributions underlying a pair of jet samples.
To minimize potential confusions, we will often use the language of quark-like and gluon-like (or ``quark'' and ``gluon'' in quotes) to refer to this operational definition.

Since the base distributions extracted from a jet observable $x$ using \demix\ are mutually irreducible, they can only agree with the MC quark and gluon jet distributions of $x$ if those are also mutually irreducible.
It was argued in \Refc{Metodiev:2018ftz} that quark--gluon mutual irreducibility is approximately satisfied for the constituent multiplicity (number of constituents) of groomed and ungroomed jets and $\nSD$~\cite{Frye:2017yrw} in proton--proton collisions, though not for other common jet observables like jet mass.
This stems from counting observables having exact quark--gluon mutual irreducibility in the high-energy limit~\cite{Frye:2017yrw}.
\Reflc{Komiske:2018vkc} further showed that constituent multiplicity is a nearly optimal classifier to separate operationally defined quark and gluon jets (see \Refc{Larkoski:2019nwj} for further developments).
We thus focus on constituent multiplicity for extracting quark and gluon fractions in our case study.

The algorithm to extract the mutually irreducible base distributions is straightforward.
Base distributions are computed from the mixture distributions via
\begin{equation}
	\label{eq:base_dists}
	\begin{aligned}
	    b_1(x) &= \frac{p_1(x) - \kappa_{1 2} \, p_2(x)}{1 - \kappa_{1 2}}\,, \\
	    b_2(x) &= \frac{p_2(x) - \kappa_{2 1} \, p_1(x)}{1 - \kappa_{2 1}}\,,
	\end{aligned}
\end{equation}
with
\begin{equation}
    \label{eq:kappa}
    \kappa_{i j} = \inf_x \frac{p_i(x)}{p_j(x)}\,.
\end{equation}
The reducibility factor $\kappa_{i j}$ is the maximum fraction of $p_j(x)$ that can be subtracted from $p_i(x)$ such that the resulting function remains positive for every $x$, and can thus be normalized to yield a proper probability distribution.
Mutual irreducibility of the $p_i$ is precisely the condition that $\kappa_{1 2} = \kappa_{2 1} = 0$.
The $\kappa_{i j}$ are directly related to the mixture fractions:
\begin{equation}
	\label{eqn:kappa}
	\kappa_{1 2} = \frac{1 - f_1}{1 - f_2}\,, \quad \kappa_{2 1} = \frac{f_2}{f_1}\,.
\end{equation}
For two analytically known mixtures of two base distributions, it is essentially trivial to compute $\kappa_{i j}$ using \cref{eq:kappa} and then to extract the mutually irreducible underlying distributions using \cref{eq:base_dists}.
As an example, two Gaussian distributions with different mean and the same standard deviation are mutually irreducible, so any two convex mixtures built from them can be demixed exactly.

\begin{figure*}[t]
\begin{subfigure}[b]{0.31\linewidth}
	\centering
	\includegraphics[width=\linewidth]{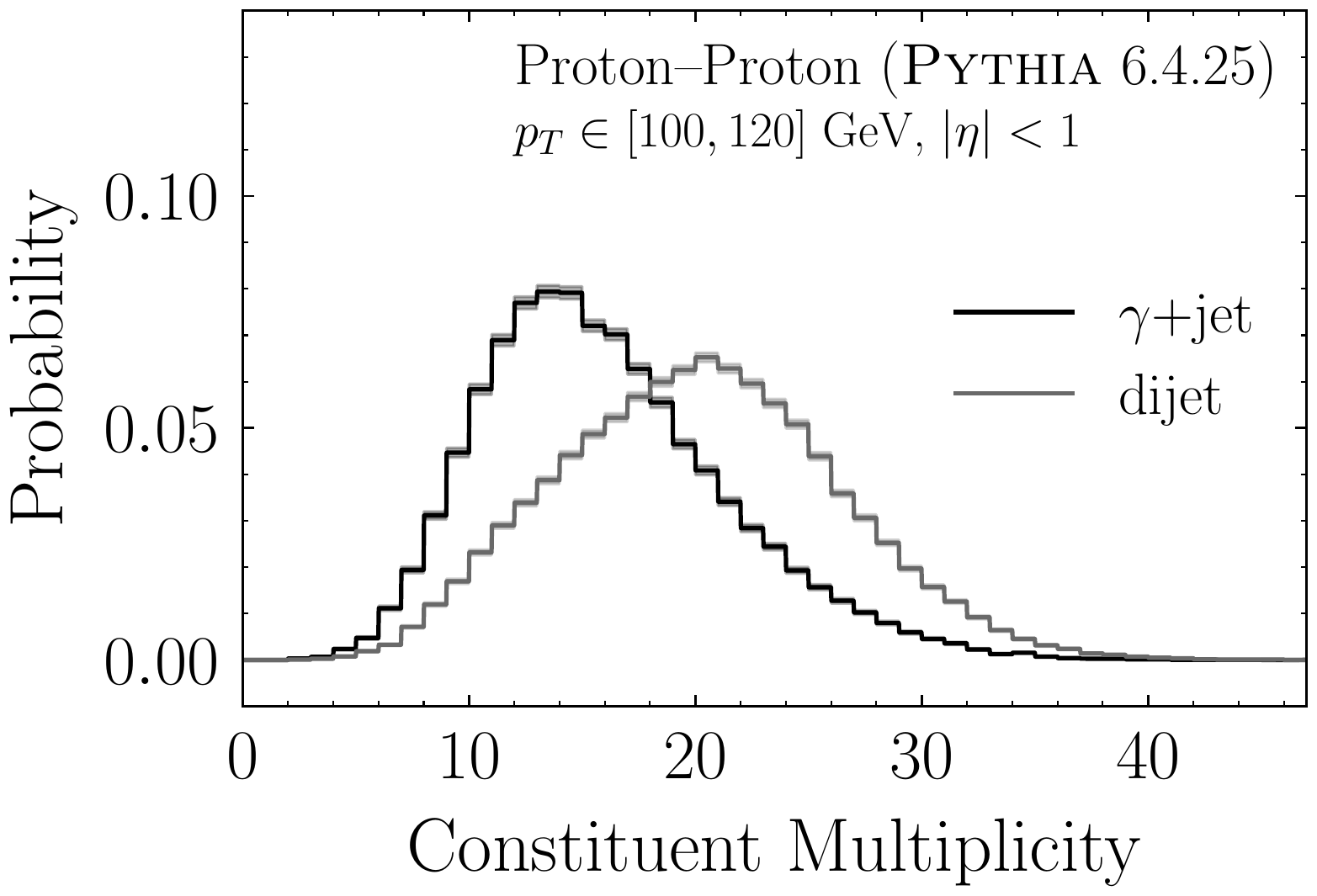}
	\caption{}
	\label{fig:dists_pp}
\end{subfigure}
\begin{subfigure}[b]{0.31\linewidth}
	\centering
	\includegraphics[width=\linewidth]{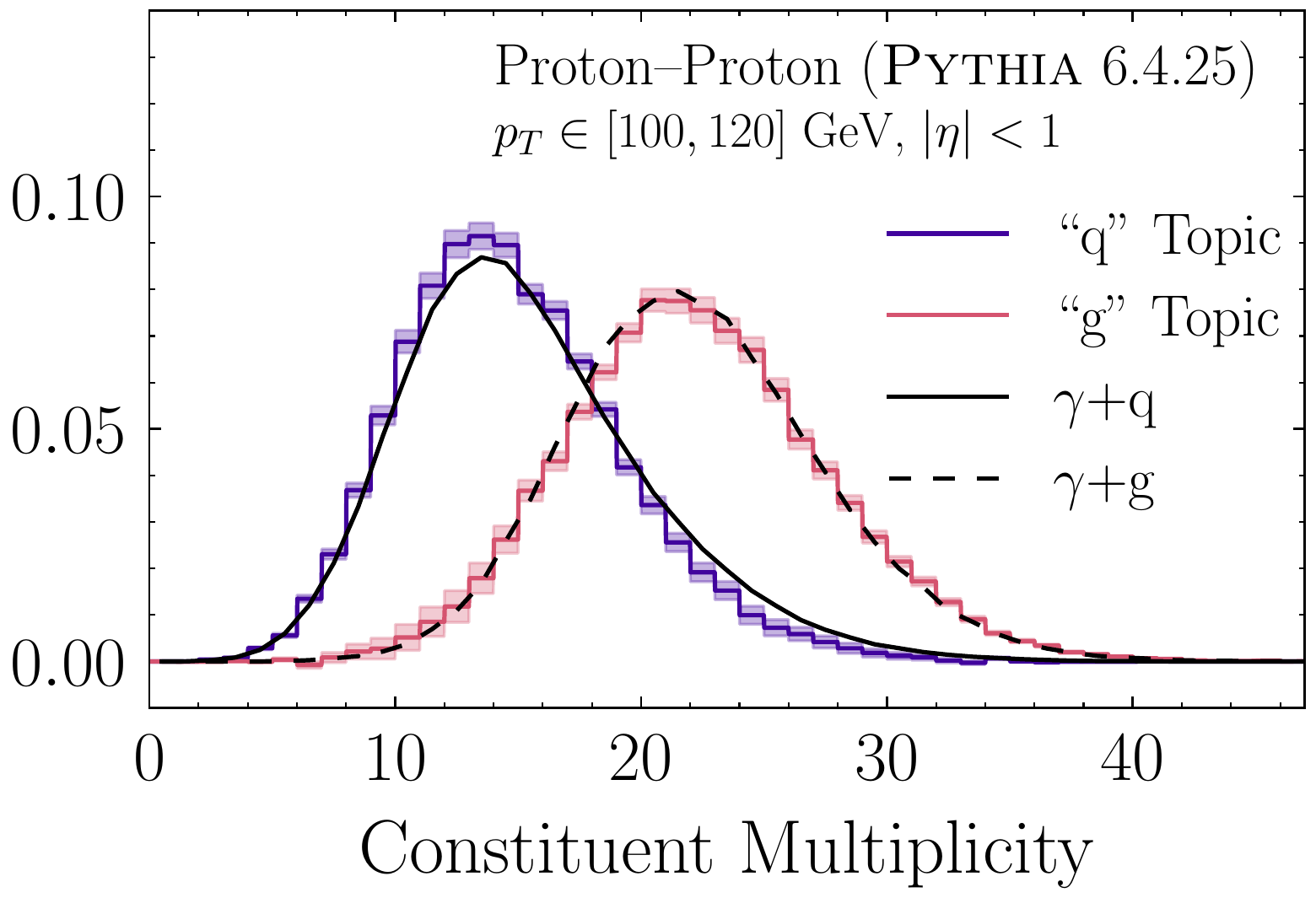}
	\caption{}
	\label{fig:topics_pp}
\end{subfigure}
\begin{subfigure}[b]{0.31\linewidth}
	\centering
	\includegraphics[width=\linewidth]{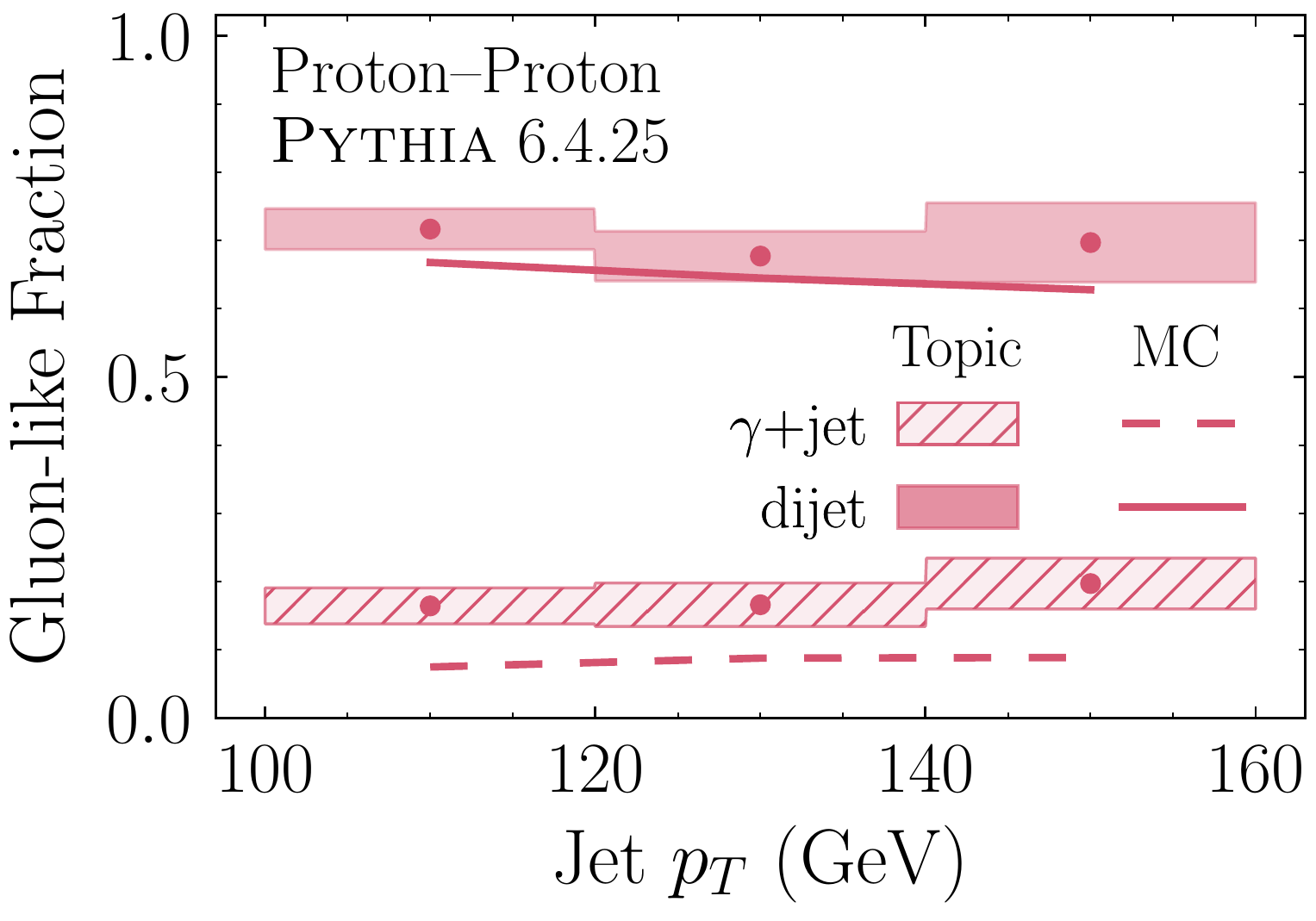}
	\caption{}
	\label{fig:fracs_pp}
\end{subfigure}
\begin{subfigure}[b]{0.31\linewidth}
	\centering
	\includegraphics[width=\linewidth]{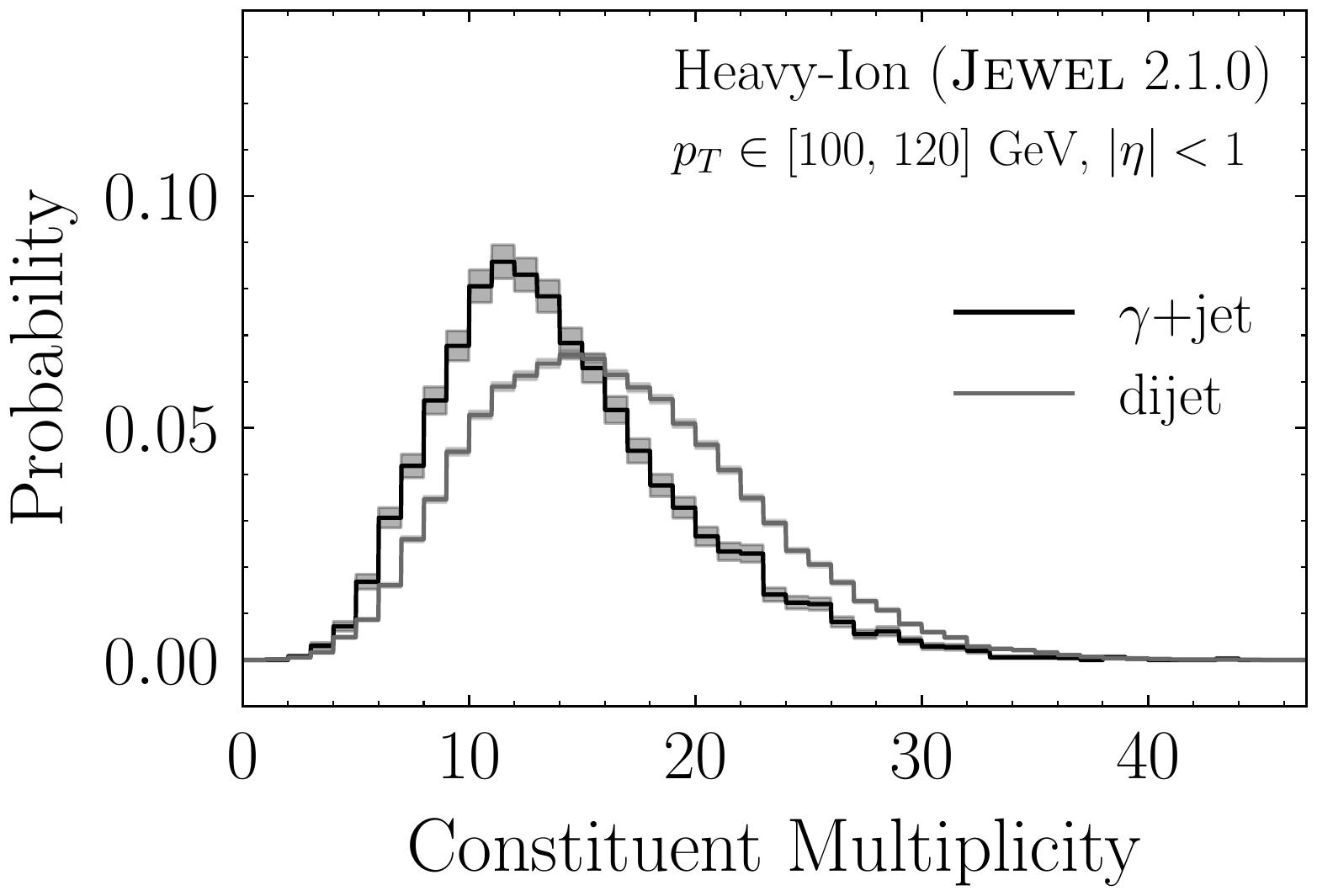}
	\caption{}
	\label{fig:dists_HI}
\end{subfigure}
\begin{subfigure}[b]{0.31\linewidth}
	\centering
	\includegraphics[width=\linewidth]{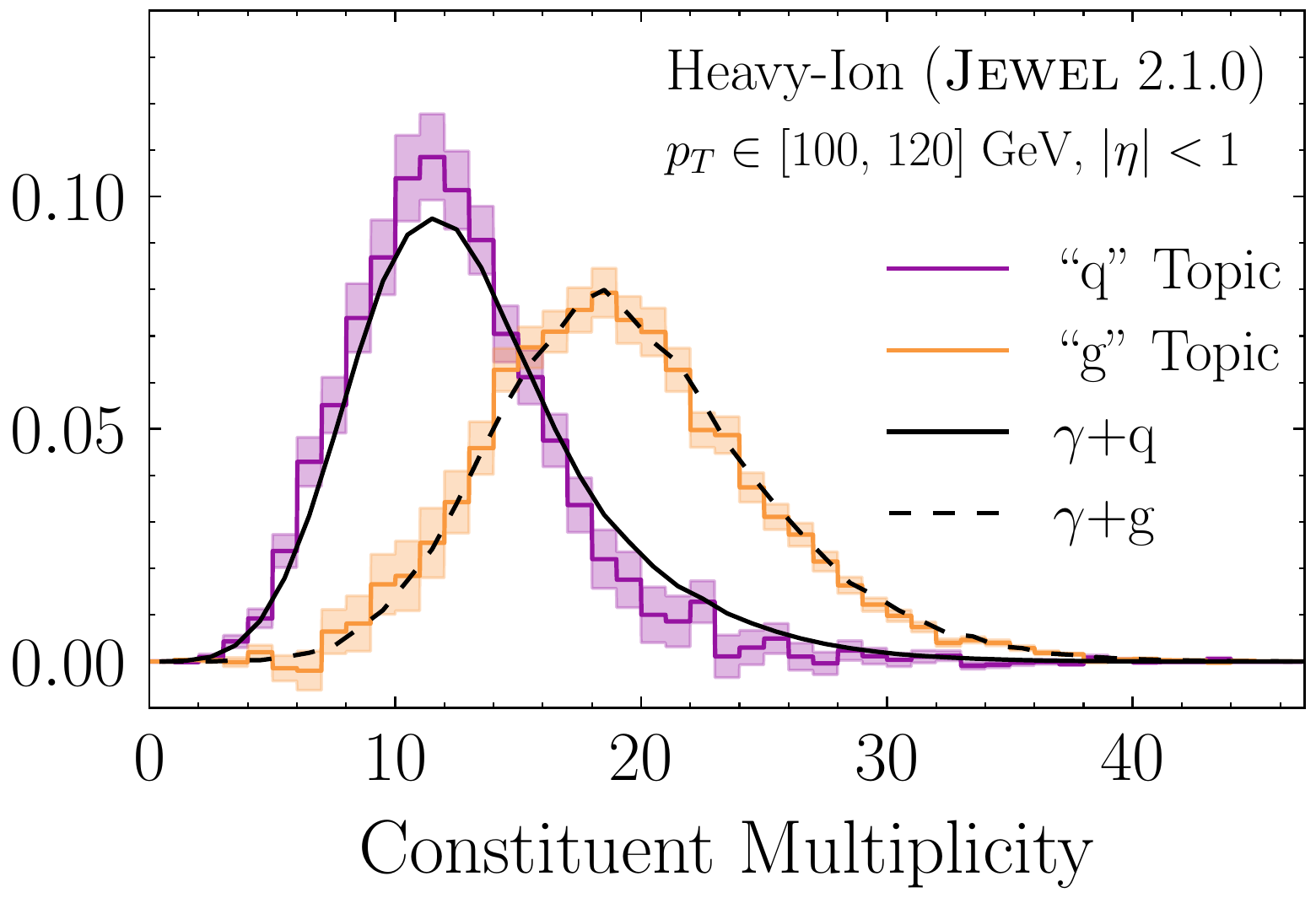}
	\caption{}
	\label{fig:topics_HI}
\end{subfigure}
\begin{subfigure}[b]{0.31\linewidth}
	\centering
	\includegraphics[width=\linewidth]{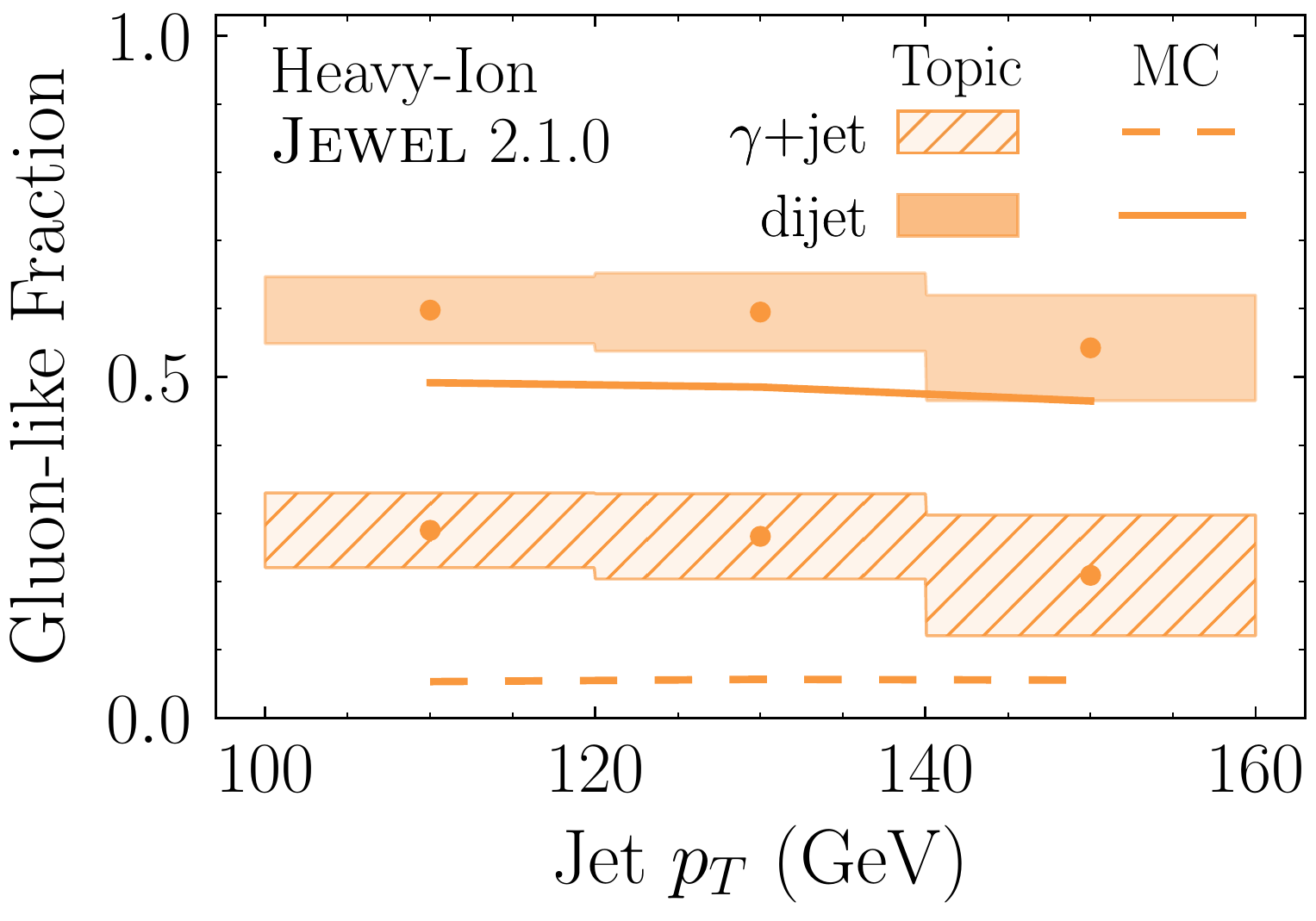}
	\caption{}
	\label{fig:fracs_HI}
\end{subfigure}
\caption{
Extracting quark-like and gluon-like jet topics from (top row) proton--proton collisions and (bottom row) heavy-ion collisions, as generated by \pythia\ and \jewel, respectively.
(a) Distributions of jet constituent multiplicity for the $\gamma+\text{jet}$ and dijet samples in proton--proton collisions.
(b) The two underlying topics extracted from these distributions using the \demix\ method (colorful bands), compared to the MC-level definition of quark- and gluon-initiated jets (black).
(c) Fractions of the gluon-like topic in the $\gamma+\text{jet}$ and dijet samples.
The corresponding results for heavy-ion collisions are shown in (d)--(f).
Possible reasons for the higher gluon-like topic fractions compared to the MC label fractions are provided in the text.}
\label{fig:topics}
\end{figure*}

When dealing with finite-sampled distributions, however, one encounters substantial technical difficulties using \cref{eq:kappa} directly.
A histogram of samples from a probability distribution $p(x)$ has a finite, discretized range of histogram bins $\{x_k\}$ at which $p(x)$ is estimated from the finite-statistics sampled distribution, $\hat{p}(x_k)$.
We need a method of defining the reducibility factors $\hat{\kappa}_{i j}$ for a pair of sampled histograms.
Naively, the infimum of \cref{eq:kappa} becomes a minimum of the ratio of the histograms over $\{x_k\}$; simply taking the minimum, however, is very sensitive to statistical fluctuations.
A more robust approach, introduced in \Refc{Komiske:2018vkc}, is to define $\hat{\kappa}_{i j}$ to be the ratio of histograms in the bin for which the ratio plus its uncertainty is minimized.
This method turns out to be insufficient to deal with the much more limited statistics we aim to utilize in this work, particularly because $\hat{\kappa}_{i j}$ is typically extracted at the low-statistics end points of the distributions.

One way to address the issue of limited statistics is by using fitting to leverage information about the interior of the distribution, where the statistics are better, to put additional constraints on the tails.
We note that the histograms shown later in \cref{fig:dists_pp,fig:dists_HI} are exceptionally well-described by a simultaneous fit to two distinct sums of a pair of skew-normal distributions $\SN(x; \mu, \sigma, s)$.
That is, they are well-described by the form
\begin{equation}
	f_N(x; \V{\alpha}_i, \V{\theta}) = \sum_{k = 1}^N \alpha_{i, k} \SN(x; \mu_k, \sigma_k, s_k)
\end{equation}
with $N = 2$.
Here $\V{\theta} = (\mu_1, \sigma_1, s_1, \dots, \mu_N, \sigma_N, s_N)$, and $\V{\alpha}_i = (\alpha_{i, 1} \dots, \alpha_{i, N})$ contains $N - 1$ independent fractions, with the $N$th fraction constrained by $\sum_{k=1}^{N} \alpha_{i,k} = 1$, for jet samples $i = 1, 2$ (dijet and $\gamma$+jet in \cref{fig:dists_pp,fig:dists_HI}).
For further generality of the functional form, we consider $N = 4$ and simultaneously fit the two input distributions to $f_4(x; \V{\alpha}_1, \V{\theta})$ and $f_4(x; \V{\alpha}_2, \V{\theta})$, respectively, with $18$ fit parameters $\V{\alpha}_1$, $\V{\alpha}_2$, and $\V{\theta}$.
To estimate the uncertainty on such fits, we use the Markov Chain Monte Carlo (MCMC) ensemble sampler \emcee~\cite{ForemanMackey:2012ig} to do posterior estimation using the likelihood function \cite{Baker:1983tu} 
\begin{equation}
	\label{eq:lnlike}
	\ln\frac{C}{p} = \sum_{i, j} n_i \left[ f(x_{i, j}; \V{\alpha}_i, \V{\theta}) - y_{i,j} + y_{i,j} \ln \frac{y_{i,j}}{f(x_{i, j}; \V{\alpha}_i, \V{\theta})}\right]\,.	
\end{equation}
%
Here, $j$ indexes the histogram bins of jet sample $i$, with the $j$th bin having constituent multiplicity $x_{i, j}$ and probability density $y_{i, j}$, and the $i$th sample having total count $n_i$.
This form assumes that the number of counts in each histogram bin, $n_{i,j} = n_i y_{i,j}$, are independently Poisson-distributed around the value $f(x_{i, j}; \V{\alpha}_i, \V{\theta})$, and estimates distributions of the parameters $\V{\alpha}_i, \V{\theta}$ for which the observed data is most likely.
Following Refs.~\cite{Baker:1983tu,Tanabashi:2018oca}, the likelihood function $p$ in \cref{eq:lnlike} is rescaled by a fit-independent constant $C$ that cancels a $\ln(n_{i,j}!)$ that arises when taking the log of the Poisson probability distribution.
We take a uniform prior on the parameters $\V{\theta}$ and $\V{\alpha}_i$ in the range $\mu_k \in [0,50]$, $\sigma_k \in [1, 15]$, $s_k \in [-20,20]$, and $\alpha_{i, k} \in [0, 1]$, and we start the MCMC walkers in a Gaussian ball of standard deviation $10\%$ around the least-squares fit parameters.
We use the distribution of fits to obtain distributions of $\hat{\kappa}_{i j}$ via \cref{eq:kappa}.
To combat finite statistics effects, we compute the infimum in \cref{eq:kappa} as a minimum of the MCMC walkers over a reduced range.
We consider only the range for which at least one input histogram is non-zero.
For each reducibility factor, we further identify whether the minimum will occur on the left or right side of the range, and truncate the opposite tail at the outermost bin that has at least 10 data points for each input histogram.
The distribution of $\hat{\kappa}_{i j}$ is used to compute a distribution of fractions, and its mean and standard deviation are used as the value and uncertainty of $\hat{\kappa}_{i j}$ used to extract the topics.

\begin{figure*}
\begin{subfigure}[b]{0.45\linewidth}
	\centering
	\includegraphics[width=\linewidth]{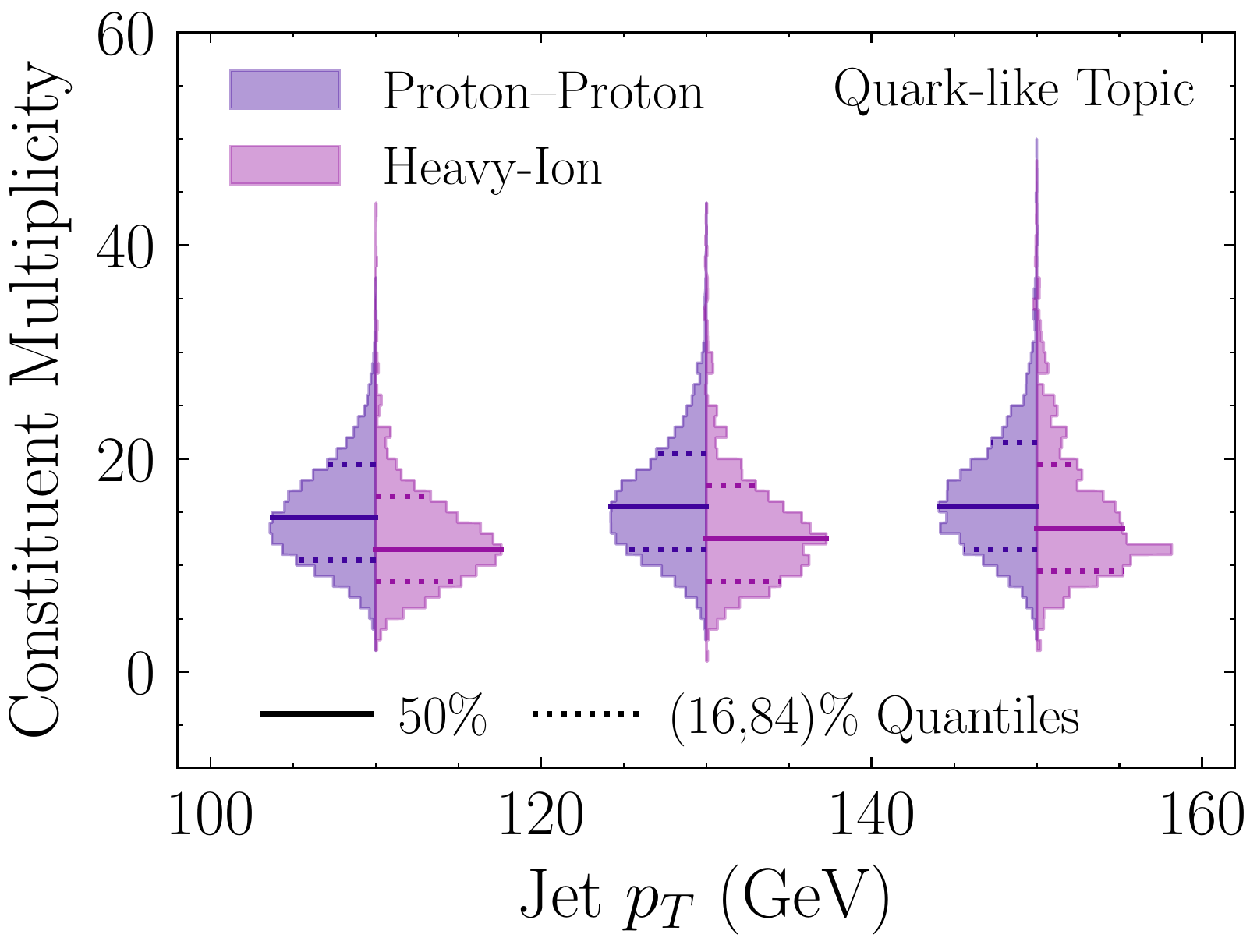}
	\caption{}
	\label{fig:dist_mod_Q}
\end{subfigure}\hspace{0.05\textwidth}
\begin{subfigure}[b]{0.45\linewidth}
	\centering
	\includegraphics[width=\linewidth]{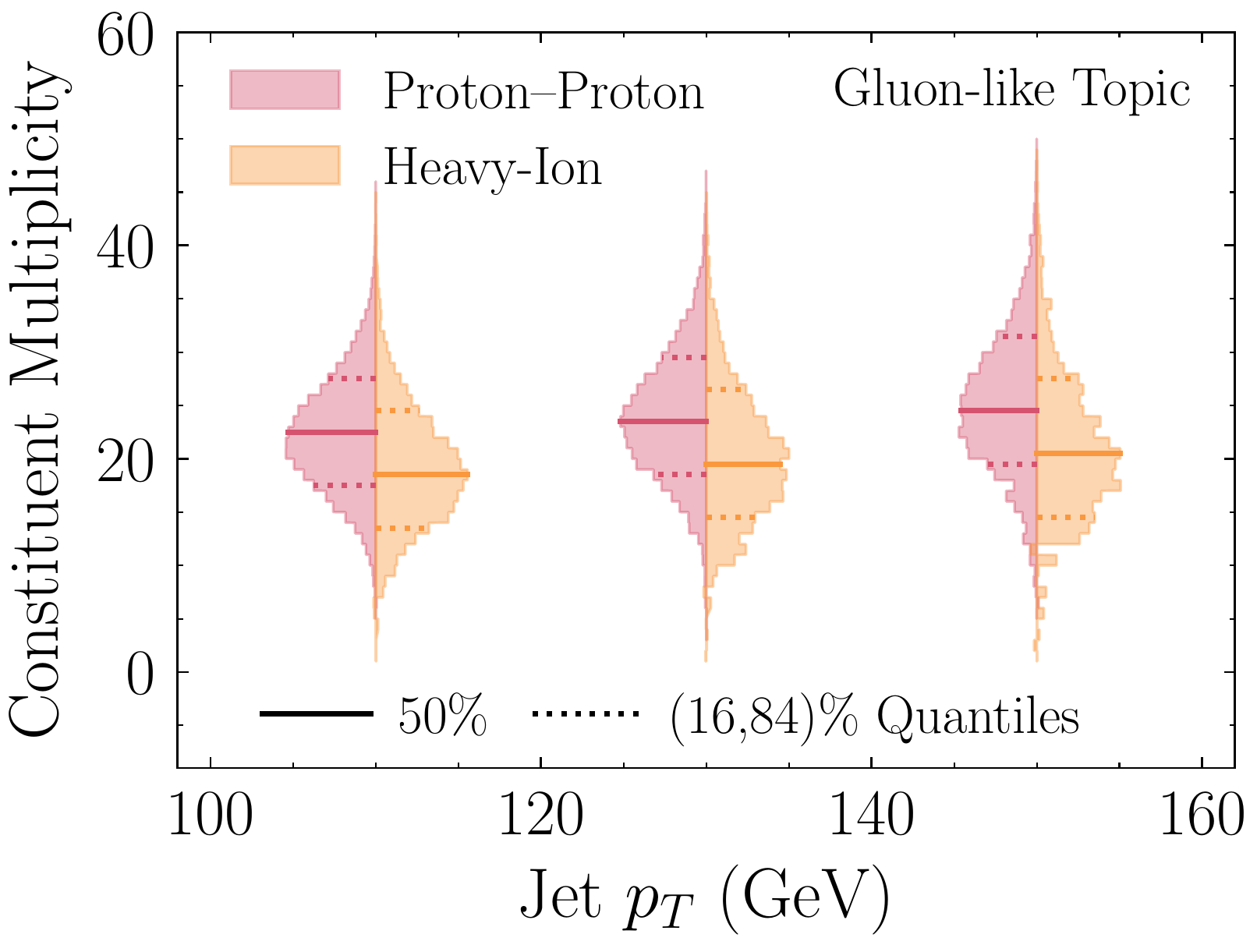}
	\caption{}
	\label{fig:dist_mod_G}
\end{subfigure}
\caption{
Constituent multiplicity distributions for (a) the quark-like topic and (b) the gluon-like topic as a function of jet $p_T$.
Each violin plot has results for both (left side) proton--proton and (right side) heavy-ion collisions, and the change between the two sides illustrates the modification of the constituent multiplicity distribution for the corresponding topic.
Horizontal lines indicate the median (solid) and $16\%$ and $84\%$ quantiles (dashed) of the multiplicity distributions.
}
\label{fig:dist_mod}
\end{figure*}

The samples for our proof-of-concept study come from the heavy-ion MC event generator \jewel\ 2.1.0~\cite{Zapp:2013vla,KunnawalkamElayavalli:2016ttl}, based on vacuum jet production in \pythia\ 6.4.25~\cite{Sjostrand:2006za}.
We consider two mixed distributions coming from photon-jet ($\gamma+\text{jet}$) production and dijet production.
For each process, we generate proton--proton and \SIrange{0}{10}{\percent} centrality heavy-ion events at $\SI{5.02}{\tera\eV}$ and reconstruct anti-$k_t$ jets using \fastjet\ 3.3.0~\cite{Cacciari:2008gp,Cacciari:2011ma} with radius parameter $R = 0.4$ within the pseudorapidity range $\abs{\eta} < 1$.
We include initial-state radiation, but do not include medium recoil effects.
(There is no underlying event model in \jewel, but we verified that similar results can be obtained after aggressively grooming jets using the Soft Drop algorithm~\cite{Larkoski:2014wba} with $z_\text{cut} = 0.5$ and $\beta = 1.5$ as in \Refc{Sirunyan:2018gct}.)
For $\gamma+\text{jet}$ events, we consider the recoiling jet with the highest transverse momentum ($p_T$), and for dijet events, we consider the two highest-$p_T$ jets.
In the case of heavy-ion collisions, we downsample our \jewel\ events to mimic the statistics that will be available with the anticipated luminosity $\int \cL \diff t = \SI{13}{\nano\barn^{-1}}$ after Run 4~\cite{Citron:2018lsq} (see Supplementary Material for details).
The equivalent statistics of our dijet sample are already less than those achievable in Run 4, but this substantially reduces the statistics of our $\gamma+\text{jet}$ sample.
We emphasize that we are only using \jewel\ for demonstration purposes, and this data-driven technique can be applied directly to experimental collider measurements for a range of jet observables beyond just multiplicity.

Starting with proton--proton collisions in the top row of \cref{fig:topics}, we show the distributions of jet constituent multiplicity for $\gamma+\text{jet}$ and dijet samples (\cref{fig:dists_pp}) and the ``quark-like'' and ``gluon-like'' topics extracted from these distributions via the data-driven method described above (\cref{fig:topics_pp}).
The corresponding heavy-ion results are shown in \cref{fig:dists_HI,fig:topics_HI}, keeping in mind that the proton--proton and heavy-ion analyses are completely independent.
The extracted topics are in good agreement with the distributions of constituent multiplicity for quark- and gluon-initiated jets as defined at the MC level.

Furthermore, we can use \cref{eqn:kappa} to extract the topic fractions, i.e., the proportions of the topics in the original input distributions.
\Cref{fig:fracs_pp,fig:fracs_HI} show the extracted fraction of the gluon-like topic in the $\gamma+\text{jet}$ and dijet samples as a function of jet $p_T$.
The gluon topic fractions are marginally higher than the MC-level fraction of gluon-initiated jets in proton--proton collisions, and more dramatically higher in heavy-ion collisions.
In interpreting these results, however, one has to be mindful of the inherent ambiguity in using MC-level information to label jets, which we explore in the Supplementary Material.
In addition, limited statistics drive the extraction of $\kappa$ from \cref{eq:kappa} into the interior of the distribution where the true minimum is not yet achieved.
In the Supplementary Material, we repeat the heavy-ion analysis using a $\gamma+\text{jet}$ sample with a factor of about $2.8$ higher luminosity.
Though the results are consistent within (large) uncertainties, the method with limited statistics will tend to overestimate the gluon-like fraction.

Even accounting for these issues, though, we find a persistently larger gluon-like fraction compared to the MC labeling, at least in the context of \jewel.
One possible explanation for this effect is that a ``quark-initiated'' jet may become more gluon-like through gluon radiation, an effect which may be enhanced by medium-induced gluon radiation in heavy-ion collisions.
For methods like this one, as well as for the method in \Refc{Sirunyan:2020qvi}, this would result in a larger fraction of jets being classified as gluon jets.
It is also possible that constituent multiplicity, though apparently nearly mutually irreducible in proton--proton collisions, may be less mutually irreducible in the presence of medium effects, so alternative observables (perhaps from machine learning~\cite{Komiske:2018vkc}) might be required.
Understanding these issues will be important for interpreting eventual LHC Run 4 data, but we emphasize that the operational definition used to define the quark-like and gluon-like topics is independent of its interpretation.

As a final proof-of-concept, in \cref{fig:dist_mod} we show the modification of the jet constituent multiplicity distributions for the quark-like (\cref{fig:dist_mod_Q}) and gluon-like (\cref{fig:dist_mod_G}) jet topics as a function of $p_T$.
To our knowledge, this represents the first fully data-driven method to separate the modification of a jet observable for ``quark'' and ``gluon'' jets.
Though we show here the modification of the constituent multiplicity distribution for clarity, we emphasize that once the topic fractions have been extracted, they can be used to extract separate quark and gluon distributions for any jet observable.
Since both jet observable distributions and the quark and gluon fractions may change between proton--proton and heavy-ion collisions, it is substantially simpler to interpret the separate modification of quark and gluon topics compared to, e.g., the modification of the dijet distribution.
Though not shown here, the method of \Refc{Brewer:2018dfs} could be used to match the proton--proton and heavy-ion jet $p_T$ quantiles and further clarify the interpretation.


In summary, we have illustrated a data-driven method to extract quark-like and gluon-like topic fractions and distributions in proton--proton and heavy-ion collisions.
Using \jewel\ samples of comparable statistics to those anticipated in Run 4 of the LHC, we have shown that these topics have a similar qualitative interpretation to the (physically ambiguous) definition of the quark and gluon jets at parton level available from MC generators.
We have further shown, as an example, the modification of the constituent multiplicity in heavy-ion collisions separately for quark and gluon jet topics.
This study offers an exciting proof-of-concept demonstration of the power of the topic modeling to interpret future heavy-ion collision data, though more quantitative studies will of course be necessary to understand the feasibility of this analysis and the best way to incorporate systematic uncertainties.
We leave a detailed study of the impact of underlying event and background subtraction, medium response, and experimental inefficiencies as important future work. 
This method is well-defined for any jet observable, and the aforementioned effects may render it important to study the performance of this analysis with additional observables beyond the constituent multiplicity.

\section{Code Availability}

Code to run this analysis on input histograms is available at \url{https://github.com/jasminebrewer/jet-topics-from-MCMC}.

\section{Acknowledgements}

The authors acknowledge valuable discussions with Liliana Apolin\'{a}rio, Raghav Kunnawalkam Elayavalli, Dhanush Anil Hangal, Patrick Komiske, Yen-Jie Lee, Eric Metodiev, Guilherme Milhano, and Krishna Rajagopal, in addition to technical assistance and feedback from Jacob Bandes-Storch, William Lewis, Lina Necib, and Anastasia Patterson.
The authors were supported by the U.S.~Department of Energy (DOE) Office of Nuclear Physics under contract DE-SC0019128 and the U.S.~Department of Energy (DOE) Office of High Energy Physics under contract DE-SC0012567, and AT was additionally supported by funding from the Tushar Shah and Sara Zion Fellowship.
%

\section{Supplementary Material}

In this supplement, we provide additional information about limited statistics effects and MC label ambiguities.

\paragraph{Monte Carlo Unweighting:}
To match the statistics available in Run 4 of the LHC, we had to downsample our heavy-ion $\gamma+\text{jet}$ events.
The expected integrated luminosity for heavy-ion collisions after Run 4 is $\cL_{A A, \text{tot}} = \int \cL \diff t = \SI{13}{\nano\barn^{-1}}$.
We should note, however, that the luminosity we are interested in is multiplied by a factor of $A^2 = 208^2$ for lead--lead collisions, $\cL_{n n, \text{tot}} = 208^2 \cL_{A A, \text{tot}}$~\cite{Roland:2014jsa}.
With the total cross-section $\sigma$ for a process, this gives the number of events $N = \sigma \cL_{n n, \text{tot}}$.

\begin{figure*}
\begin{subfigure}[b]{0.31\linewidth}
	\centering
	\includegraphics[width=\linewidth]{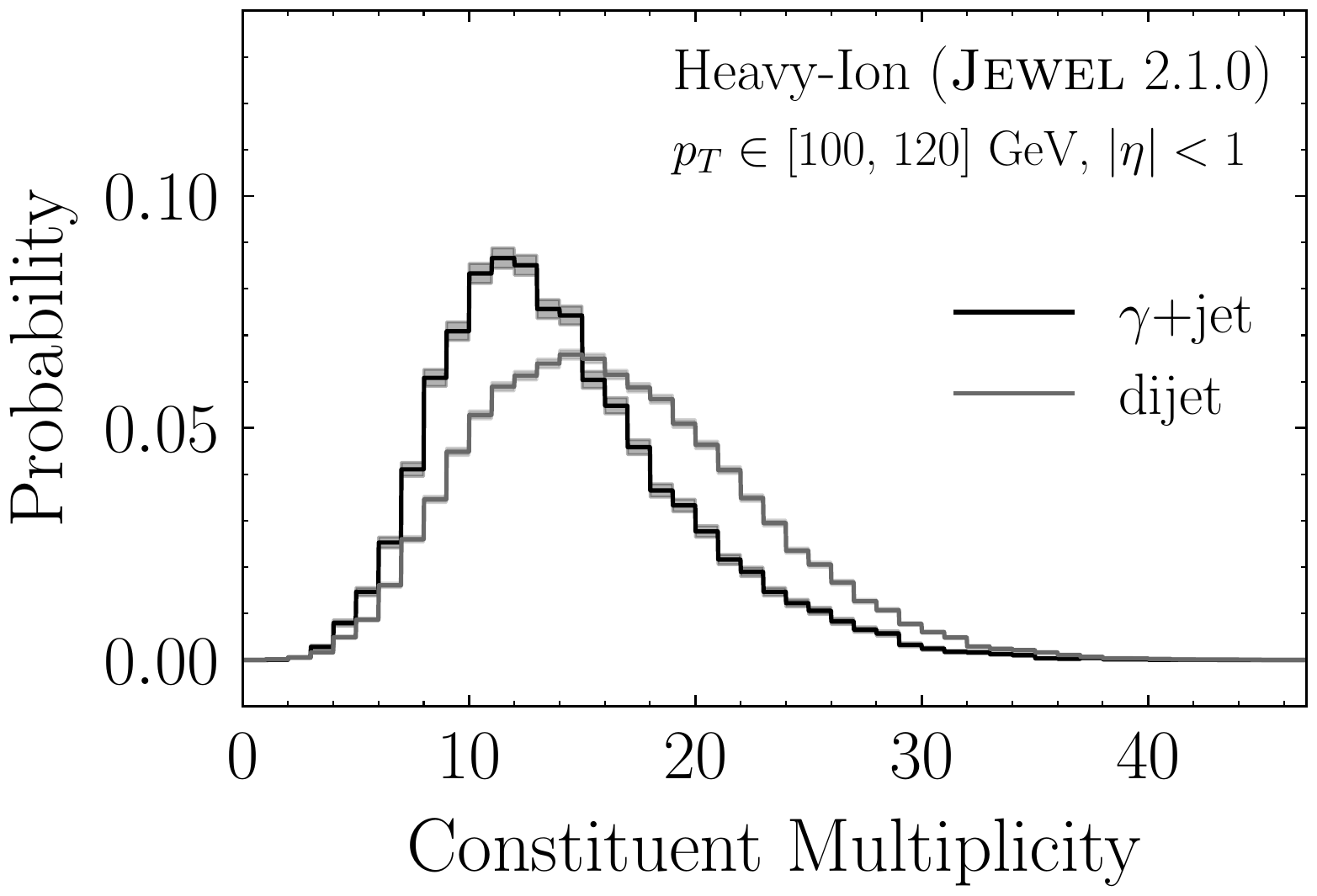}
	\caption{}
	\label{fig:dists_allstats}
\end{subfigure}
\begin{subfigure}[b]{0.31\linewidth}
	\centering
	\includegraphics[width=\linewidth]{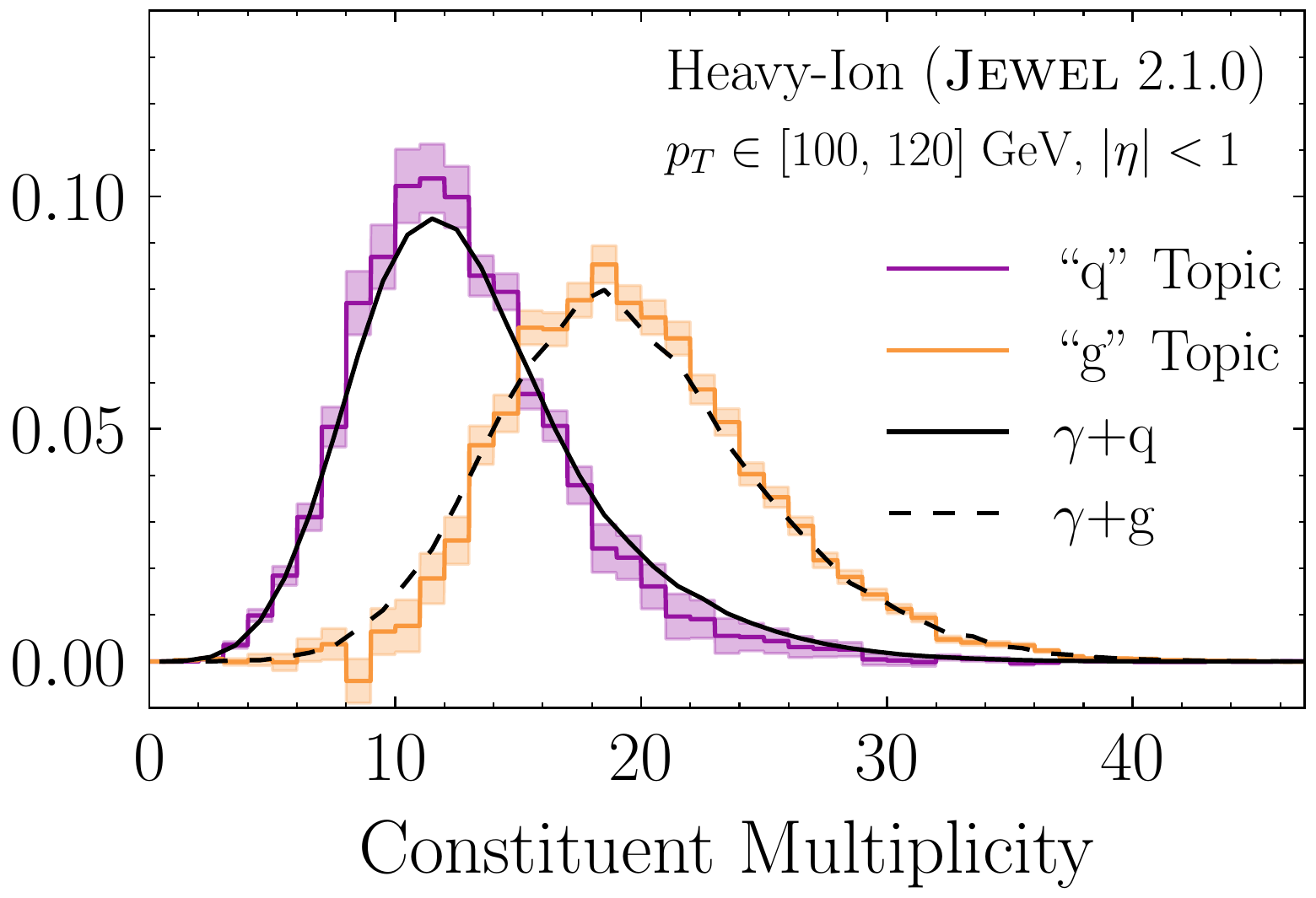}
	\caption{}
	\label{fig:topics_allstats}
\end{subfigure}
\begin{subfigure}[b]{0.31\linewidth}
	\centering
	\includegraphics[width=\linewidth]{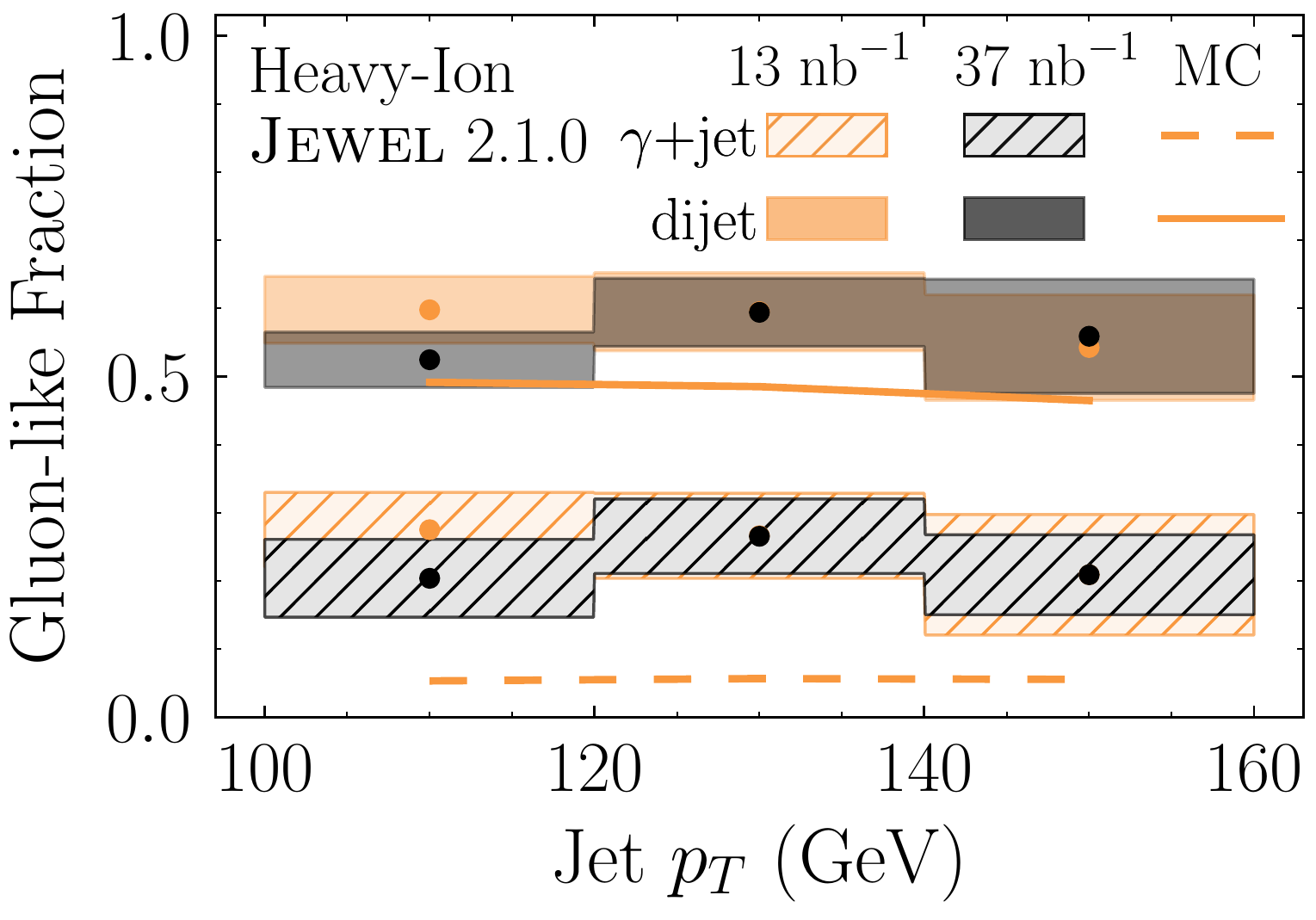}
	\caption{}
	\label{fig:fracs_allstats}
\end{subfigure}
\caption{Same as the bottom row of \cref{fig:topics}, but using a factor of about $2.8$ higher $\gamma+\text{jet}$ statistics than expected after LHC Run 4.
With more events, the agreement of the distribution of the quark-like topic (purple) relative to the MC-level definition (black) is somewhat improved compared to \cref{fig:topics_HI}, though the gluon-like topic fraction remains high compared to the MC label fraction.
}
\label{fig:allstats}
\end{figure*}

Since our \jewel\ sample is based on weighted events, we perform a standard unweighting procedure to simulate the statistics possible after Run 4.
If we have $N_w$ weighted events, the unweighting procedure probabilistically chooses some subset of these events to act as unweighted events.
To obtain an expected number of kept events $N$, the $j$th event is (independently) kept with probability $p_j = N \frac{w_j}{w_\text{tot}}$, where $w_\text{tot} = \sum_j w_j$.
Note that this puts a constraint on the required size of $N_w$ to be able to unweight it to a sample of size $N$, namely that $1 \ge p_\text{max} = N \frac{w_\text{max}}{w_\text{tot}}$, with $w_\text{max}$ being the maximum value of $w_j$.
We unweight our proton--proton sample per $p_T$ bin by fixing $p_\text{max}=1$ in each bin and downsampling to a sample of expected size $N$.
After this unweighting we have $63603$ and $42466$ jets in dijet and $\gamma$+jet samples, respectively, with $p_T \in [100,120] \text{ GeV}$, $25046$ and $18488$ with $p_T \in [120,140] \text{ GeV}$, and $13707$ and $9260$ with $p_T \in [140,160] \text{ GeV}$.
For clarity, we downsample our heavy-ion results uniformly across all $p_T$ bins so that the resulting statistics correspond to a fixed luminosity of $\SI{13}{\nano\barn^{-1}}$ (and later $\SI{37}{\nano\barn^{-1}}$ for \cref{fig:allstats}).
Our $\SI{13}{\nano\barn^{-1}}$ $\gamma$+jet sample has $6453$, $2825$, and $1436$ jets in $p_T$ ranges $[100,120] \text{ GeV}$, $[120,140] \text{ GeV}$, and $[140,160] \text{ GeV}$, respectively.
Our dijet sample has lower luminosity with $71098$, $32318$, and $15747$ jets, respectively, in the same $p_T$ bins.

\begin{figure*}
\begin{subfigure}[b]{0.4\linewidth}
	\centering
	\includegraphics[width=\linewidth]{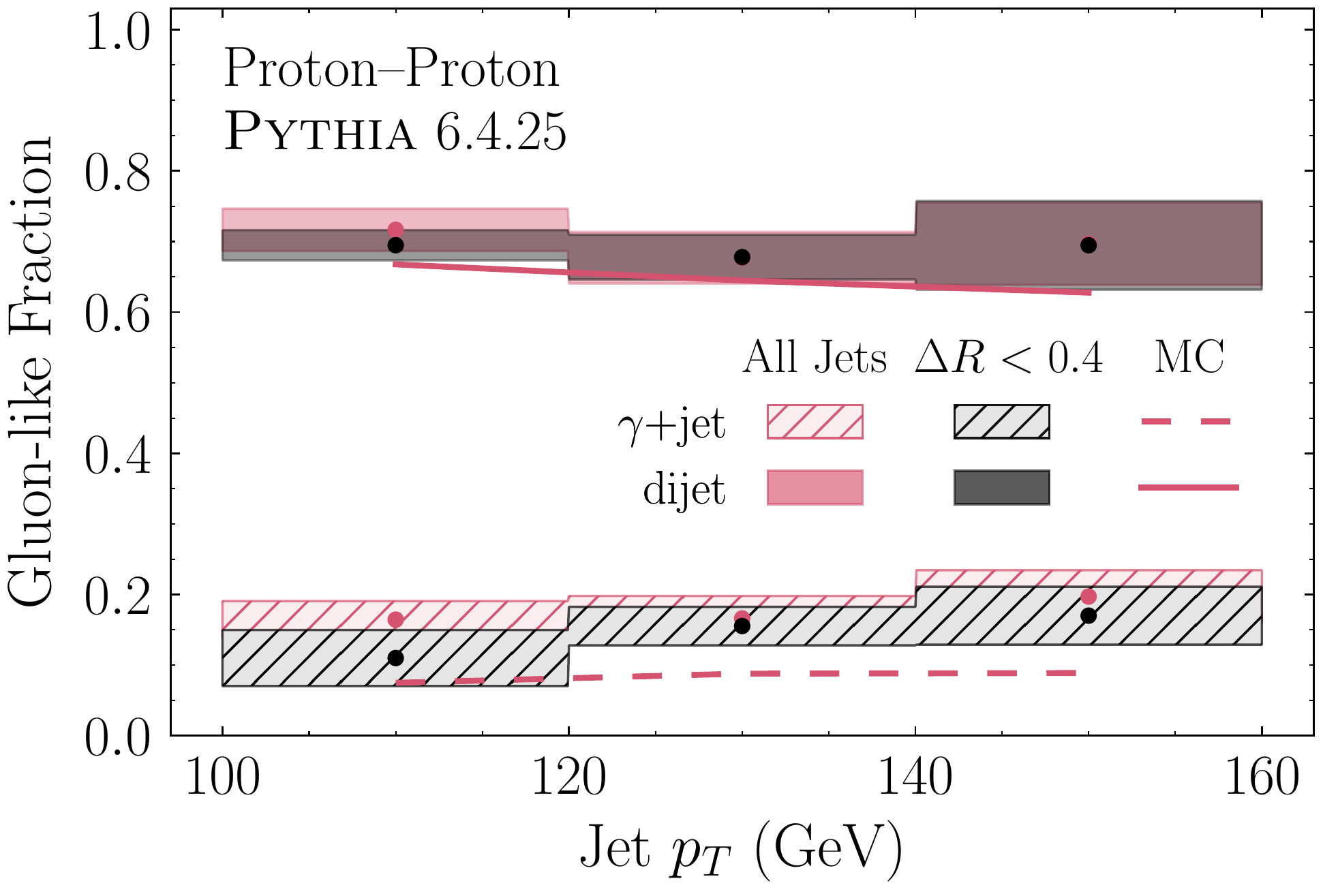}
	\caption{}
	\label{fig:fracs_pp_dR}
\end{subfigure}\hspace{0.05\textwidth}
\begin{subfigure}[b]{0.4\linewidth}
	\centering
	\includegraphics[width=\linewidth]{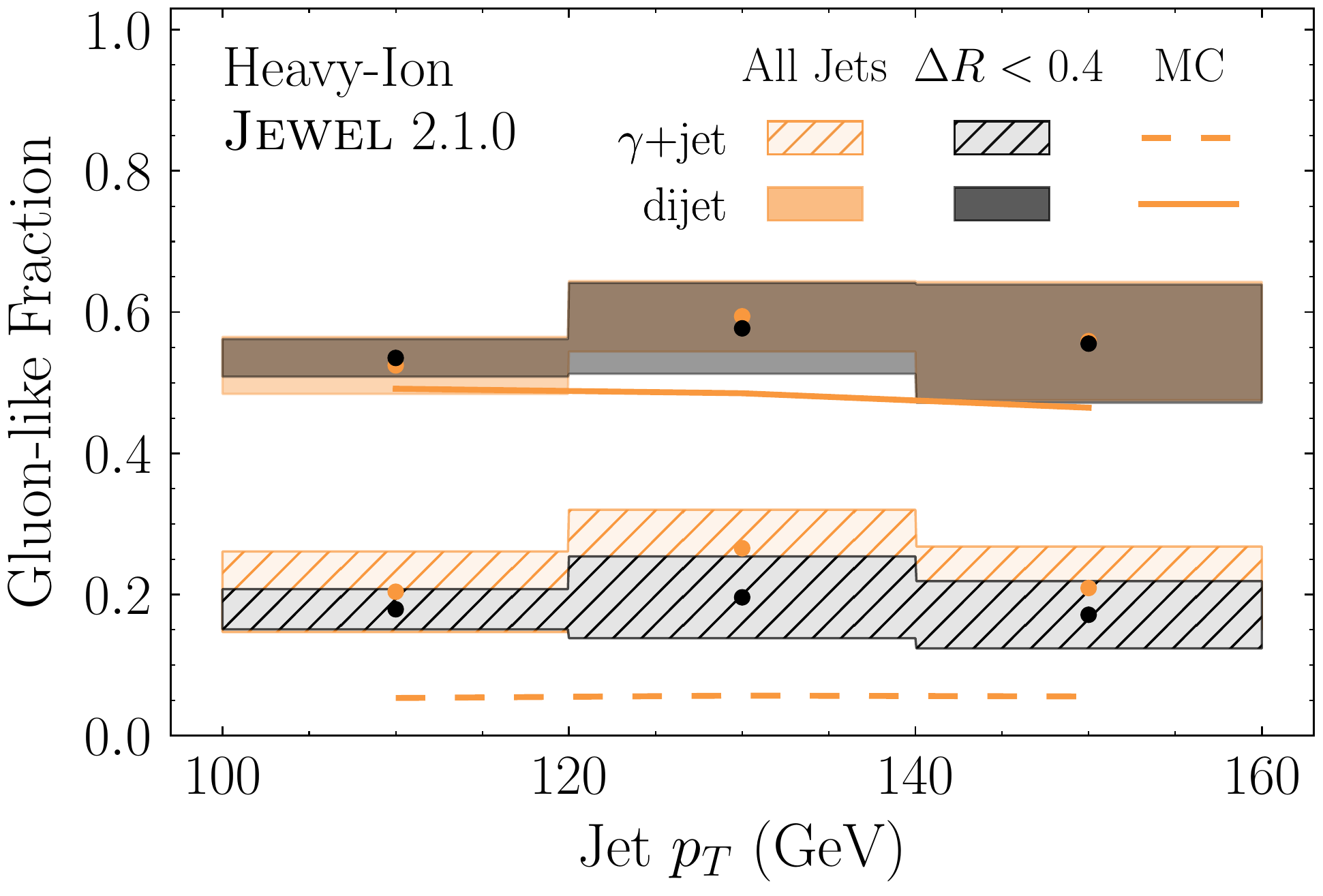}
	\caption{}
	\label{fig:fracs_HI_dR}
\end{subfigure}
\caption{Gluon-like topic fractions for (a) proton--proton collisions and (b) heavy-ion collisions, as a function of jet $p_T$, for the $\gamma+\text{jet}$ (hatched) and dijet (solid) samples.
Colorful bands are the extracted gluon-like topic fractions from the full $\gamma+\text{jet}$ and dijet distributions, as already shown in \cref{fig:fracs_pp,fig:fracs_HI}.
Black bands show for comparison the gluon-like topic fractions extracted from (unphysically) restricted $\gamma+\text{jet}$ and dijet samples including only jets whose axis is within $\Delta R = 0.4$ of a parton in the hard scattering matrix element.
}
\label{fig:fractions_dR}
\end{figure*}

\paragraph{Limited Statistics Effects:}
While the proton--proton results shown in the main body of the text are relatively robust to statistical effects, the heavy-ion results are impacted by the lower statistics expected for LHC Run 4.
This particularly affects the $\gamma+\text{jet}$ sample, where the statistics only enable a rough estimate of the constituent multiplicity distribution.
In \cref{fig:allstats}, we show the result of topic extraction after increasing the $\gamma+\text{jet}$ sample to have a factor of about $2.8$ more events than expected in LHC Run 4.
With limited statistics, the method tends to overestimate the gluon-like fraction, though the results are consistent within uncertainties.
With the higher statistics, the gluon-like topic fractions still remain somewhat higher than the MC-level fractions, which could simply be due to residual limited statistics effects.
As mentioned in the main text, this effect could alternatively have a physical origin and might arise from ``quark-initiated'' jets becoming more gluon-like via gluon radiation, larger deviations from mutual irreducibility for constituent multiplicity in the presence of medium effects, or ambiguities in the MC flavor labeling.

\paragraph{Monte Carlo Label Ambiguities:}
It is important to remember that MC-level information is inherently ambiguous.
In the main text, we defined the MC quark- and gluon-initiated jet distributions ($\gamma+\text{quark}$ and $\gamma+\text{gluon}$ in \cref{fig:topics}) to include only those from jets whose axis is within $\Delta R = \sqrt{\Delta \eta^2 + \Delta \phi^2} = 0.4$ from a quark or gluon in the initial hard scattering matrix element.
This requirement is only satisfied, however, by $87\%$ and $90\%$ of the jets in our heavy-ion sample of $\gamma+\text{jet}$ and dijets, respectively, with $p_T \in [100,120]$~GeV.
In the analogous proton--proton samples, it is satisfied for $94\%$ and $95\%$ of the jets in $\gamma+\text{jet}$ and dijet samples, respectively.
(These percentages can be further increased by requiring the two jets, or the photon and jet, be nearly back-to-back in azimuthal angle.)
Particularly for our heavy-ion results, this implies that the $\gamma+\text{jet}$ and dijet samples are not direct combinations of the MC-level quark and gluon jet distributions.

Thus, one might wonder whether part of the mismatch between the gluon-like fractions seen in \cref{fig:fracs_HI} could be due to the fundamental ambiguity in jet flavor labeling.
In \cref{fig:fractions_dR}, we study this by (unphysically) restricting the $\gamma+\text{jet}$ and dijet distributions to those jets whose axis is within $\Delta R=0.4$ of an initiating parton, and then using these restricted samples to perform the topic extraction.
This has relatively little impact on the gluon-like topic fraction, though it tends to make it somewhat more similar to the MC-level gluon-initiated jet fraction.
%

\bibliography{jet-topics}

\end{document}